# Pseudometrically Constrained Centroidal Voronoi Tessellations: Generating uniform antipodally symmetric points on the unit sphere with a novel acceleration strategy and its applications to Diffusion and 3D radial MRI

Cheng Guan Koay

Department of Medical Physics
University of Wisconsin School of Medicine and Public Health
Madison, WI 53705

*Corresponding author:*
*Cheng Guan Koay, PhD*
*Department of Medical Physics*
*University of Wisconsin School of Medicine and Public Health*
*1161 Wisconsin Institutes for Medical Research (WIMR)*
*1111 Highland Avenue*
*Madison, WI 53705*
*E-mail: cgkoay@wisc.edu*

**ABSTRACT**
**Purpose:**

The purpose of this work is to investigate the hypothesis that uniform sampling measurements that are endowed with antipodal symmetry play an important role when the raw data and image data are related through the Fourier relationship as in q-space diffusion MRI and 3D radial MRI. Currently, it is extremely challenging to generate large uniform antipodally symmetric point sets suitable for 3D radial MRI. A novel approach is proposed to solve this important and long-standing problem.

**Methods:** The proposed method is based upon constrained centroidal Voronoi tessellations of the upper hemisphere with a novel pseudometric. Geometrically intuitive approach to tessellating the upper hemisphere is also proposed.

**Results:** The average time complexity of the proposed centroidal tessellations was shown to be effectively linear. For small sample size, the proposed method was comparable to the state-of-the-art iterative method in terms of the uniformity. For large sample size, in which the state-of-the-art method is infeasible, the reconstructed images from the proposed method has less streak and ringing artifact as compared to those of the commonly used methods.

**Conclusion:** This work solved a long-standing problem on generating uniform sampling points for 3D radial MRI.

## INTRODUCTION

A centroidal Voronoi tessellation (2) is a Voronoi tessellation (3) in which the center of mass (or centroid) of the Voronoi region is also its generator and it has been found useful in many applications, from analysis of cellular pattern (4), neuronal density (5) and territorial behavior of animals(6) in biological sciences to optimal data acquisition, data quantization, compression and clustering (7) in the engineering and physical sciences. However, it remains a challenge to generate these tessellations via Lloyd's algorithm even though Lloyd's algorithm has been found to be very robust (8).

The main computational bottleneck in the computation of the centroidal Voronoi tessellations has been the reconstruction step in which the Voronoi tessellations are reconstructed anew from the newly computed centroids at each iteration. From the conventional point of view, if the time complexity of the Voronoi tessellations of a specific manifold, e.g., plane, sphere or other esoteric surfaces, is O(n log n), then the time complexity of the centroidal Voronoi tessellations on this manifold must be O(m n log n), i.e., invoking the O(n log n) algorithm m number times. Note that m is the number of iterations needed to reach convergence and n is the number of generators.

We have discovered strong heuristic strategies to reduce the average time complexity of the problem from O(m n log n) to O(m n) by using less optimal, i.e., O($n^2$), but simpler algorithm innovatively and will present these strategies in the context of generating uniform antipodally symmetric points on the sphere. The approach used in this work for accelerating the centroidal Voronoi tessellations is completely novel, representing a clear departure from the current paradigms.

The problem of generating uniform points on the sphere was proposed by J.J. Thomson (9) more than a century ago and has spurred many theoretical and computational investigations, and applications (1,10-15). A variant of the Thomson problem is that of generating uniform antipodally symmetric points on the sphere (16); the resultant point set plays a particular important role in MRI, which will be discussed next. It is interesting to note that the problem of generating nearly uniform antipodally symmetric points on the sphere via a deterministic method was solved only recently (13,14).

Conceptually, diffusion MRI (17) and 3D radial MRI (18) share many similar features. The most notable feature relevant to this work is the data redundancy associated with symmetries inherent in each of these imaging techniques. Specifically, antipodal symmetry of the diffusion propagator (19) and Hermitian symmetry due to the real-valuedness of the object in the image domain influence how data are acquired in diffusion MRI and 3D radial MRI, respectively. Hermitian symmetry and antipodal symmetry are closely related to each other and both symmetries are defined through parity transformation or spatial inversion with Hermitian symmetry having an additional operation, which is complex conjugation. Specifically, a real-valued function $f$ possesses antipodal symmetry if $f(\mathbf{x}) = f(-\mathbf{x})$ and a complex-valued function $g$ possesses Hermitian symmetry if it is a Hermitian function, i.e., $g(\mathbf{x}) = \overline{g(-\mathbf{x})}$ where complex conjugation is denoted by the overhead bar. Due to these symmetries, it is sufficient to sample only half of the space, e.g., q-space in diffusion MRI or k-space in MRI. However, certain constraints on the imaging gradients (20) force the sampling trajectory in 3D radial MRI to take the usual diametrical line or a curve consists of two straight radial lines adjoined with a slight non-smooth bend at the center of k-space (21); the former sampling strategy cannot take advantage of the data redundancy because sampling along a diametrical line through k-space only is equivalent to acquiring repeated measurements and not new measurements and the latter sampling strategy is more desirable with the potential of acquiring new and different points in k-space.

The use of uniformity and antipodal symmetry in sampling scheme was advocated by Jones (16) in diffusion tensor imaging on the basis that the diffusion tensor is symmetric. Here, we argue that uniformity and antipodal symmetry should be incorporated into sampling design in diffusion MRI (q-space formalism) and 3D radial MRI (k-space formalism) because both the q-space and k-space possess Hermitian symmetry. Uniform sampling measurements endowed with antipodal symmetry can maximize sampling coverage and at the same time take advantage of the inherent symmetry of the imaging systems. It is to the best of our knowledge that none of the works in 3D radial MRI (18,21-28) used antipodally symmetric point sets to acquire k-space data. In fact, many studies (22,23,25,27) used the method of Wong and Roos (15) or the generalized spiral scheme (10,11). The point sets generated from any of the methods, i.e., generalized spiral scheme, the method of Wong and Roos and the method of Bauer (12), have the spiral pattern although each of these methods was formulated and derived differently. Antipodally symmetric point sets have not been used in any 3D radial MRI study may be due to the following reasons: the lack of a robust iterative method or a deterministic method capable of generating large number of antipodally symmetric points on the sphere, and the availability of other deterministic methods (1,11,12,15,29) for generating non-antipodally symmetric point sets. Ad hoc strategy of collecting half of the non-antipodally symmetric points on one of the hemispheres is usually adopted to come up with an antipodally symmetric point set. We should note that the diffusion MRI sampling method based on vertices of a multifold-tessellated icosahedral hemisphere was used by Tuch et al. (30) but this method is inflexible and limited in that it cannot generate point set of any size except certain "magic" numbers, e.g., 6, 26, 46, 91, 126, 196, 246, 341, 406, 526 and so on. Note that fivefold-tessellated icosahedral hemisphere produces 126 points on the upper hemisphere, see Tuch (30). Due to the above mentioned limitation, this sampling method used by Tuch et al. may not be very appealing to the practitioners of 3D radial MRI.

Even with the advent of the deterministic method capable of generating nearly uniform antipodally symmetric points on the sphere (13,14), it is still extremely challenging to generate highly uniform antipodally symmetric point set for use in 3D radial MRI through iterative function minimization even when the initial solution is taken from the deterministic methods. The number of points (on the sphere) required in 3D radial MRI and diffusion MRI is in the thousands (4000 to 20000 or more) and in the hundreds (6 to 500), respectively. Therefore, the number of parameters in spherical coordinate representation to be optimized are twice as many (8000 to 40000) and optimization methods that rely on approximate or exact Hessian or its inverse are completely infeasible. While some of the iteratively optimized antipodally symmetric point sets (with sample size in the hundreds) have been tabulated and available for public use (31), the lack of highly uniform and antipodally symmetric point sets applicable to 3D radial MRI motivates the present investigation.

Here, we propose a constrained centroidal Voronoi tessellation endowed with a pseudometric to robustly and efficiently generating uniform antipodally symmetric point sets on the unit sphere. The uniformity of the point sets generated from the proposed method is shown to be comparable to the state-of-the-art iterative method (31) for small sample size relevant to diffusion MRI but not to 3D radial MRI. For large point set with sample size in the range of thousands, which is relevant to 3D radial MRI and other applications (32), the state-of-the-art method is completely infeasible but the proposed approach remains feasible. Further, we used our recently developed three-dimensional analytical MRI phantom in the Fourier domain (33), which was based upon the three-dimensional version of the famous Shepp-Logan phantom (34,35) in the image domain, to compare and contrast the qualitative features of the reconstructed images obtained through the proposed method, the method of Wong and Roos, the generalized scheme and the tessellated-icosahedral of the upper hemisphere (the method used by Tuch et al.). We found that the reconstructed image from the proposed method has less streak and ringing artifacts as compared to those of other methods.

## METHODS

A centroidal Voronoi tessellation (2) is a Voronoi tessellation in which the center of mass (or centroid) of the Voronoi region is also its generator and the Voronoi regions can be prescribed with a density function. To the best of our knowledge, a centroidal Voronoi tessellation capable of generating uniform antipodally symmetric points on the unit sphere has never been proposed. Instead of dealing with a density function that is constant and invariant with respect to spatial inversion or antipodally symmetry, it is more convenient and efficient to introduce a novel pseudometric in the centroidal Voronoi tessellation so as to make generating uniform antipodally symmetric points on the unit sphere possible.

The spherical Voronoi regions, $\{V_i\}_{i=1}^{N}$, on the upper hemisphere are characterized by the following properties:

- No two distinct regions share the same point. That is, the intersection between any two distinct regions is empty, $V_i \cap V_j = \varnothing$ with $i \neq j$. Points on the boundary between any two regions belong to the closure of these regions, which is denoted by an overhead bar, e.g., $\overline{V_i}$.

- The union of all the closures of the spherical Voronoi regions covers the upper hemisphere, denoted by $S_+^2$.

- $V_i = \{\mathbf{x} \in S_+^2 \mid d(\mathbf{x}, \mathbf{g}_i) < d(\mathbf{x}, \mathbf{g}_j) \quad \text{for} \quad j = 1, ..., N \quad \text{and} \quad j \neq i\}$  [1]

Each of the unit vectors, $\{\mathbf{g}_i\}_{i=1}^{N}$, on $S_+^2$ is called the generator of its respective Voronoi region. The pseudometric, $d$, will be defined later.

The center of mass of each Voronoi region does not necessarily coincide with the generator of that region. An iterative method such as the Lloyd algorithm (8,36) may be used to make the generators from each successive iterations closer and closer to the centers of mass of their respective regions. The resultant tessellations are known as the centroidal Voronoi tessellations (2).

The center of mass of a spherical Voronoi region, $V_i$, can be expressed as:

$$\hat{\mathbf{g}}_i = \frac{\int \mathbf{v} d\sigma_i}{\left\| \int \mathbf{v} d\sigma_i \right\|} \; . \tag{2}$$

Note that $\sigma_i$ is the spherical surface of $V_i$, $\mathbf{v}$ represents unit vector normal to the spherical surface element $d\sigma_i$ and $\|.\|$ denotes the Euclidean norm, which is needed to ensure $\hat{\mathbf{g}}_i$ is unit length. In practice, the computation of $\hat{\mathbf{g}}_i$ is based on the discretized version of Eq.[2]. Specifically, it is obtained through the sum of the products between the area of the spherical triangle formed by the generator and each pair of consecutive vertices at the boundary surrounding the generator in counterclockwise order. We further note that the density function, which usually appears as a factor in the integrand of Eq.[2], is taken to be a unit constant function in order to ensure uniformity of the generators.

The distance measure or metric, $d(\cdot,\cdot)$, used in this work is a novel extension of the modified electrostatic potential energy term studied in Ref.(37). For completeness, we will introduce the notion of real and virtual points for manipulating the antipodally symmetric point set suggested in Ref.(37). Due to the constraint of antipodal symmetry, we classify points as real and their corresponding antipodal points as virtual. If we have $N$ real points (on the upper hemisphere), denoted by unit vectors $\mathbf{r}_i$ with $i = 1, \cdots, N$, then the total electrostatic energy for the complete configuration (37) of $2N$ points of both real and virtual points on the whole sphere is given by:

$$\varphi \;=\; \frac{N}{2} \;+\; 2 \sum_{i=1}^{N-1} \sum_{j=i+1}^{N} \left( \frac{1}{r_{ij}} + \frac{1}{\sqrt{4 - r_{ij}^2}} \right) \tag{3}$$

with $r_{ij} \equiv \left\| \mathbf{r}_i - \mathbf{r}_j \right\|$. Note that Eq.[3] is expressed solely in terms of real points. If we define

$$S(\mathbf{r}_i, \mathbf{r}_j) \equiv \frac{1}{r_{ij}} + \frac{1}{\sqrt{4 - r_{ij}^2}}$$ then $S(\mathbf{r}_i, \mathbf{r}_j)$ may be thought of as a reciprocal metric (or reciprocal of

the distance measure) between two real points.). Here, we exploit this reciprocal metric fully by

defining our pseudometric, $d(\cdot, \cdot)$, as

$$d(\mathbf{r}_i, \mathbf{r}_j) = 1 / S(\mathbf{r}_i, \mathbf{r}_j). \tag{4}$$

Note that $d(\mathbf{r}_j, \mathbf{r}_j) = 0 = d(\mathbf{r}_j, -\mathbf{r}_j)$, hence the term pseudometric. Please refer to Appendix B

for the proof that $d(\cdot, \cdot)$ is indeed a pseudometric.

We should note that implementation of spherical Voronoi tessellations is not a trivial computational task (38,39) and our approach is different from the existing methods,

Refs.(38,39). To compare and contrast our proposed approach, we will first present an outline of

the Lloyd's algorithm below:

---

Pseudometrically Constrained Centroidal Voronoi Tessellation:
Let $n$ be the number of desired points on the upper hemisphere.

Step 1. Deterministically generate $2n$ highly uniform points on the unit sphere
via Analytically Exact Spiral Scheme(1) and select those points that are
on the upper hemisphere as the generators.
Step 2. Construct the spherical Voronoi regions of the upper hemisphere with the
chosen generators.
Step 3. Compute the normalized centroids using the discretized version of Eq.[2].
Step 4. Adopt the normalized centroids as the generators.
Step 5. Iterate Steps 2, 3, 4 until convergence it reached.

We would like to point out that time complexity of Step 2 above is O(n log n), see Ref.(39). Therefore, the time complexity of the centroidal Voronoi tessellation is O(m n log n) where m is the number of iterations needed to reach convergence.

Our approach to generating the Voronoi tessellations of the upper hemisphere is based on the following steps and observations:

1. The Voronoi region of a generator can be found by first identifying surrounding generators (approximately neighbors of neighbors) within a spherical cap with a prescribed radius, which depends on the total number of generators, away from the generator of interest, see Figure 1A. This step can always be done because the initial point set obtained from the Analytically Exact Spiral Scheme(1) is already reasonably uniform. When the total number of generators is large, the number of surrounding generators remain constant and is around 20. The time complexity of this step for all the generators is $O(n^2)$.

2. The generator and its surrounding generators are rotated in such a way that the generator is situated along the z-axis, see Figure 1A. The time complexity of this step for all the generators is O(n).

3. To compute the vertices of the Voronoi region, the smallest convex region formed by the surrounding generators and around the generator is needed, see Figure 1B. Finding this smallest convex region is equivalent to finding the boundary points of the convex hull of the surrounding generators after these generators have been stereographically projected (40) onto the x-y plane, see Figure 1C. The validity of this step hinges on the angle preserving (or conformal) property of steoreographic projection (41). The boundary points of the convex hull of a set of points on the plane is then computed through the Graham's scan, (42). The time complexity of Graham's scan is O(k log k) and k is the number of points in the convex hull, i.e., the number of surrounding generators, which is about 20. Therefore, the time complexity of this step for all the generators is O(n) because k is a constant.

4. The vertices of the Voronoi region, the area of spherical triangles formed by the Voronoi vertices and the generator can be computed in O(n) for all the generators.

The steps proposed above are intuitive and geometrically motivated but we should note that Step 1 in our proposed approach is $O(n^2)$, which is the main bottleneck of our approach if this step were to be invoked at every iteration. In what follows, we will describe a simple way to avoid tessellating at every iteration and provide empirical evidence that our approach is effectively O(n) for generating the centroidal Voronoi tessellations. We can compute the distance between a generator at the current iteration and the same generator from the previous iteration. This information has already been used to determine convergence, i.e., Ref.(8). The novelty of our proposal is in using this information to compute the cumulative sum of Euclidean distances made by each generators. If the maximum value of these cumulative sums is less than some prescribed value, see Appendix A, then Step 1 will not be invoked. Otherwise, Step 1 will be invoked and the cumulative sum is reset to zero. When Step 1 is not invoked, the connectivity network between a generator and its surrounding generators will not be altered but the coordinates of the generator and its surrounding generators will likely be different from one iteration to the next.

Here, we discuss other important implementation details. Note that some of the vertices around the equator may be located at the lower hemisphere and the resultant centroids may also be on the lower hemisphere. Therefore, the centroids or generators should be reoriented onto $S_+^2$ at each iteration. Finally, we note that the convergence of the proposed algorithm is based on that of local deviation as described in Ref.(8).

## RESULTS

We implemented the proposed pseudometrically constrained centroidal Voronoi tessellation in Java. We conducted four tests to illustrate the proposed method. The first two tests were run on a machine with an Intel® Core™ i7 CPU at 1.73 GHz and with 8 GB of RAM and the last two tests were run on a different machine, Four X 8-Core 2.3GHz AMD Opteron Processors (6134) with 128GB of RAM.

The first test is to show that the performance of the point sets generated by the proposed method is of comparable quality in terms of modified electrostatic energy to that of the state-of-the-art iterative method and to show that the proposed centroidal Voronoi algorithm is O(m n) through empirical analysis of the execution time per iteration as a function of number of generators (or points on the upper hemisphere), which shows a linear trend, see Figure 2.

Figure 2A shows the percent of relative error in terms of the modified electrostatic energy of initial deterministic generators, which were obtained our recent analytically exact spiral scheme(1) and of the final generators obtained by the proposed method with respect to the state-of-the-art method. At the level of percent of relative error achieved by the proposed method, the uniformity of the points between the proposed method and the state-of-the-art method is indistinguishable visually or quantitatively. However, we observed that the point sets generated centroidal Voronoi tessellations has consistently shown to be higher in modified electrostatic energy than those obtained through iterative scheme. We believe this observation has to do with the intrinsic property of centroidal Voronoi tessellations, that is the fundamental centroidal constraint itself. The centroidal constraint forces generators to be in geometrically frustrated positions even though the effect of the centroidal constraint on geometric frustration is very slight.

Figure 2B shows the performance of the proposed method in terms of execution time (in seconds) per iteration of the proposed method as a function of the number of points on the upper hemisphere. It is clear that the execution time per iteration has a linear trend with respect to N.

Figure 2C shows the frequency of invocation of Step 1, which is an $O(N^2)$ algorithm, as a function of N, number of points on the upper hemisphere and Figure 2D shows the total number of iterations as a function of N and the instances of invocation of Step1 that occurred within these iterations are shown in red. Note that the highest peak is at (N=324, 4161 iterations).

The second test is to give an example of a point set with sample size large enough that the state-of-the-art method is currently not feasible. In this example, we used sample size of 888. Figure 3A shows the initial generators and their Voronoi regions as obtained the analytically exact spiral scheme(1). Figures 3B and 3C shows the centroids and centroidal Voronoi regions on the upper and the lower hemispheres, respectively. Figures 3D shows the centroids on the whole sphere by combining Figures 3B and 3C.

The third test is intended to show the feasibility of the proposed method in generating point set of sample size large enough for 3D MRI applications and beyond. Figure 4 shows only the Voronoi regions on the sphere with 16000 generators on the upper hemisphere. It took 10.38 minutes to generate the centroidal Voronoi tessellations (at the tolerance level of $10^{-12}$) on the machine with Four X 8-Core 2.3GHz AMD Opteron Processors (6134).

The final test is intended to show the qualitative features of the reconstructed images of the three-dimensional analytical MRI phantom (33) obtained from the proposed method, the method of Wong and Roos, the generalized spiral scheme and the tessellated icosahedral of the upper hemisphere. Note that the reconstruction algorithm is based upon the following non-uniform discrete Fourier transform for the image domain signal at $\mathbf{r}$ :

$$I(\mathbf{r}) = \sum_{i=1}^{m} S(\mathbf{k}_i) \, |\, \mathbf{k}_i \,|^2 \exp(2\pi i \, \mathbf{k}_i \cdot \mathbf{r}) \, ,$$

where $S(\mathbf{k}_i)$ is the signal value of the three-dimensional Shepp-Logan phantom evaluated analytically at $\mathbf{k}_i$, see (33). The term, $|\, \mathbf{k}_i \,|^2$, came about from the spherical coordinate transformation in the Fourier expansion. Two target matrix sizes of the reconstructed imaging volume were chosen and they were $128 \times 128 \times 128$ (standard resolution) and $256 \times 256 \times 256$ (high resolution). We have $m = 526 \times 64 = 33664$, $k_{\max} = 32$ (arbitrary units) and $\Delta k = 0.5$ (arbitrary units) at the standard resolution and $m = 526 \times 128 = 67328$, $k_{\max} = 64$ (arbitrary units) and $\Delta k = 0.5$ (arbitrary units) at the high resolution. Figure 5 shows the reconstructed images generated from the different schemes studied in this work at the standard resolution (the first two rows) and at the high resolution (the last two rows). At the standard resolution of 128x128x128 with z = -0.181, the images generated by the method of Wong and Roos and the generalized spiral scheme have more noticeable ringing artifacts around the bright region than that of the proposed method. The image generated by the tessellated icosahedral scheme has more streak artifacts in the dark regions than that of the proposed method. At the same resolution with z = 0.228, the images generated from these methods have more ringing artifacts than that of the proposed method. Similar patterns of artifacts showed up more noticeably at a higher resolution of 256x256x256. At this resolution, the images generated from the proposed method have less ringing artifacts than those from other methods.

## DISCUSSION

Even though our recent works (13,14) were the first to solve the problem of deterministically distributing antipodally symmetric points on the unit sphere, the generated points are structurally very regular and are prone to causing degeneracy in triangulation (43), i.e., when four points rather than three sit on the same circumcircle, and hence, point sets from these works could not be used as the initial generators for the proposed method. The point sets generated from the analytically exact spiral scheme did not have the above mentioned defect and contribute immensely to the feasibility of the present approach by providing highly efficient technique for generating nearly uniform points on the sphere and half of which are then used as the initial generators in the present approach.

The novelty of this work lies in the realization that the reciprocal of the reciprocal metric can be treated as a pseudometric in the antipodally symmetric space, in the extension of the centroidal Voronoi tessellations to the antipodally symmetric space, in the heuristic strategies suggested above for accelerating centroidal Voronoi tessellations without tessellating at every iteration, and in the utilization of the proposed constrained centroidal Voronoi tessellations to generate uniform antipodally symmetric points on the sphere. It should be clear that the pseudometric used in this work can also be used in any K-mean clustering algorithm of discrete data that are endowed with antipodal symmetry. The results showed that antipodally symmetric uniform sampling strategy does play a positive role in image quality and such a strategy should be incorporated into sampling design considerations of any 3D radial MRI study.

One of the most exciting results of this work is the heuristic strategies used to accelerate centroidal Voronoi tessellations. This result highlights the emergence of strong heuristics in which traditional optimal algorithm that is invoked iteratively may not outperform less optimal algorithm that is invoked sparingly in an innovative way to accomplish the same computational task. For small sample size, the point sets generated from the proposed method are comparable to those generated from the state-of-the-art method in all the cases tested. One of the important highlights of this work is that the proposed method is capable of generating large uniform point sets relevant to 3D radial MRI applications; the state-of-the-art iterative method is completely infeasible in this case particularly when the desired number of points is in the thousands.

We must note that the reconstruction algorithm adopted in this work is for testing purposes as it is computational expensive but it serves the present purpose very well because whatever the differences in the reconstructed images these differences have to come from uniformity or lack thereof of each of the methods studied in this work. It is not fortuitous that the reconstructed image from the proposed method has less streak and ringing artifacts as compared to those of other methods studied in this work.

Uniform antipodally symmetric point sets have continued to play a major role in MRI acquisition design (14,16,23,37,44-49) and are beginning to make an impact in other fields (32). We hope this work contributes to further advances to these scientific endeavors beyond diffusion and 3D radial MRI.

## Acknowledgment

The author dedicates this work to Lean Choo Khoo. He would like to thank Drs. M. Elizabeth Meyerand, Charles A. Mistretta and Peter J. Basser for the encouragement. Thanks go to Dr. Orhan Unal, Director of Medical Physics Computing, for making available the computing resources needed for this work and for providing the specifications of the Opteron Processors. The author would like to thank Drs. Steven R. Kecskemeti, Kevin M Johnson, and Michael A. Speidel for helpful discussion on image reconstruction and image quality assessment. This work was supported in part by the National Institutes of Health IRCMH090912. The software related to this work will be made freely available for research use at URL: http://sites.google.com/site/hispeedpackets/ .

**Appendix A**

In this appendix, we present a method, which is an special case of our previous work(50), to compute the radius of a spherical cap so as to cover sufficient number of surrounding generators for the purposes of computing Voronoi region. We also mention the criterion adopted to decide whether to invoke Step 1 or otherwise.

If the center of a circle of radius $r$ on a plane that is tangent to a unit sphere at point (0,0,1) or on the z-axis, then its inverse Gnomonic projection is a spherical cap, see Figure 2A of Ref.(50). Since the area of the upper hemisphere of unit radius is $2\pi$, the area associated with a generator can be approximated by $2\pi/N$ where $N$ is the total number of generators on the upper hemisphere. This approximate works well for large $N$, i.e., 15 and above. The unnormalized areal measure, Eq.(33) of Ref.(50), of a spherical cap with area equals to $2\pi/N$ is related to $r$ by the following expression:

$2\pi/N = 2\pi(1 - \frac{1}{\sqrt{1+r^2}})$.

Therefore, $r = \sqrt{\left(\frac{N}{N-1}\right)^2 - 1}$ .

The angle subtended by an longest arc passing through the interior of the spherical cap is given by:

$\theta = 2\tan^{-1}(r)$.

This angle is also the approximate spherical distance between two generators. Therefore, any generators that are within $5\theta/2$ in spherical distance away for the generator of interest will be classified as surrounding generators.

Previously, we mentioned that if the maximum value of the cumulative sums of distances made by generators in successive iterations is less than some prescribed threshold value, which is a small fraction of the prescribed radius in Step 1, then Step 1 will not be invoked. The prescribed value is based on Euclidean distance for simplicity as the local convergence criterion is also based on Euclidean distance. The prescribed threshold value is given

$C\sqrt{2(1-\cos(\tan^{-1}(r)))}$ where $C$ is a positive real number less than unity. The specific value of $C$ adopted in this study was $C = 0.15$. Note that if $C$ assumes too high a value, one may run into the problem in which the initial surrounding generators may not be the current surrounding generators, which may cause the error in Voronoi tessellations. If $C$ assumes too low a value, Step1 may be invoked many more times than necessary.

**Appendix B: The proposed pseudometric for the unit sphere**

The distance measure, $d(\cdot,\cdot)$, proposed in this work is a novel extension of the modified electrostatic potential energy term studied in (37). For the sake of completeness, the notion of real and virtual points for manipulating the antipodally symmetric point set as suggested in (37) will be reintroduced here.

Due to the constraint of antipodal symmetry, we classify points as real and their corresponding antipodal points as virtual. If we have $N$ real points (say on the upper hemisphere), denoted by unit vectors $\mathbf{r}_i$ with $i = 1, \cdots, N$, then the total electrostatic energy for the whole configuration (37) of $2N$ points of both real and virtual points on the whole sphere is given by:

$$\varphi = \frac{N}{2} + 2 \sum_{i=1}^{N-1} \sum_{j=i+1}^{N} \left( \frac{1}{r_{ij}} + \frac{1}{\sqrt{4 - r_{ij}^2}} \right),$$   [B1]

with

$$r_{ij} \equiv \begin{cases} \left\| \mathbf{r}_i - \mathbf{r}_j \right\| & if \quad \arccos(\mathbf{r}_i \cdot \mathbf{r}_j) \leq \pi/2 \\ \left\| \mathbf{r}_i + \mathbf{r}_j \right\| & if \quad \arccos(\mathbf{r}_i \cdot \mathbf{r}_j) > \pi/2 \end{cases}.$$   [B2]

We should note that we could have defined $r_{ij}$ to be simply $\left\| \mathbf{r}_i - \mathbf{r}_j \right\|$, which is in fact the actual definition used in the implementation of the proposed algorithm, and the expression in Eq.[B1] would still be valid but the definition of $r_{ij}$ in Eq.[B2] would bring much clarity to the proof later. Note that Eq.[B1] is expressed solely in terms of real points. The interaction term in Eq.[B1],

$$I(\mathbf{r}_i, \mathbf{r}_j) \equiv \frac{1}{r_{ij}} + \frac{1}{\sqrt{4 - r_{ij}^2}}$$   [B3]

was first postulated in (37) as a reciprocal metric (or reciprocal of the distance measure) between two real points.). Here, we prove that $m(\cdot,\cdot)$ given by

$$m(\mathbf{r}_i, \mathbf{r}_j) \equiv 1 / I(\mathbf{r}_i, \mathbf{r}_j) = \frac{1}{\frac{1}{r_{ij}} + \frac{1}{\sqrt{4 - r_{ij}^2}}} \qquad [B4]$$

is indeed a pseudometric.

Before we begin the proof to show that $m(\cdot, \cdot)$ is a pseudometric on the unit sphere, we first list all the necessary information about (1) the pseudometric conditions on the unit sphere, (2) concave functions, (3) sub-additive functions and (4) a well known property about the vertices of any spherical triangle on the unit sphere that is endowed with antipodal symmetry.

Definition 2.1. Let $S^2$ be the three-dimensional unit sphere. A pseudometric on $S^2$ is a mapping, $d : S^2 \times S^2 \to [0, 2]$, such that

(a) $d(\mathbf{r}_i, \mathbf{r}_j) \geq 0$,

(b) $d(\mathbf{r}_i, \mathbf{r}_j) = d(\mathbf{r}_j, \mathbf{r}_i)$,

(c) $\mathbf{r}_i = \mathbf{r}_j$ implies $d(\mathbf{r}_i, \mathbf{r}_j) = 0$, and

(d) $d(\mathbf{r}_i, \mathbf{r}_j) \leq d(\mathbf{r}_i, \mathbf{r}_k) + d(\mathbf{r}_j, \mathbf{r}_k)$.

Note that $\mathbf{r}_i$, $\mathbf{r}_j$, and $\mathbf{r}_k$ are unit vectors. Note further that the $d(\cdot, \cdot)$ is considered a metric if it satisfies the above conditions and also satisfies the converse of (c), i.e., $d(\mathbf{r}_i, \mathbf{r}_j) = 0$ implies $\mathbf{r}_i = \mathbf{r}_j$. In brief, a metric is automatically a pseudometric but the converse is not generally true. It is not hard to see that $m(\cdot, \cdot)$ satisfies the first three conditions trivially by virtue of Eq.[B2]. The proof will focus on showing that $m(\cdot, \cdot)$ also satisfies the last condition, i.e., the triangle inequality.

Definition 2.2. A function $g$ is concave on an interval $[a, b]$ if, for any $x$ and $y$ in $[a, b]$ and for any $\alpha \in [0, 1]$, we have

$$g(\alpha x + (1-\alpha) y) \geq \alpha g(x) + (1-\alpha) g(y).$$  [B5]

Lemma 2.1. If a function $g$ is concave on an interval $[0, b]$ for some $b > 0$ and $g(0) = 0$, then $g$ is subadditive, i.e., $g(x+y) \leq g(x) + g(y)$ for any $x, y \geq 0$.

Proof. By setting $y = 0$ and invoking the fact that $g$ is concave and $g(0) = 0$, Eq.[B5] reduces to

$$g(\alpha x) \geq \alpha g(x).$$  [B6]

Therefore,

$$g(x+y) = \frac{x}{x+y} g(x+y) + \frac{y}{x+y} g(x+y) \leq g(\frac{x}{x+y}(x+y)) + g(\frac{y}{x+y}(x+y)) = g(x) + g(y).$$

The following lemma is well known and it is stated for completeness.

Lemma 2.2. Any three points, $\mathbf{r}_i, \mathbf{r}_j$, and $\mathbf{r}_k$, on $S^2$ that are endowed with antipodal symmetry, two of the eight triplets, $(\pm\mathbf{r}_i, \pm\mathbf{r}_j, \pm\mathbf{r}_k)$, have vertices bounded by a spherical octant.

Consequently, any one of these two triplets, say, $\widetilde{\mathbf{r}}_i, \widetilde{\mathbf{r}}_j$, and $\widetilde{\mathbf{r}}_k$, satisfies the following properties:

$$\arccos((\widetilde{\mathbf{r}}_k - \widetilde{\mathbf{r}}_i) \cdot (\widetilde{\mathbf{r}}_j - \widetilde{\mathbf{r}}_i)) \leq \pi/2,$$

$$\arccos((\widetilde{\mathbf{r}}_k - \widetilde{\mathbf{r}}_j) \cdot (\widetilde{\mathbf{r}}_i - \widetilde{\mathbf{r}}_j)) \leq \pi/2, \text{ and}$$

$$\arccos((\widetilde{\mathbf{r}}_j - \widetilde{\mathbf{r}}_k) \cdot (\widetilde{\mathbf{r}}_i - \widetilde{\mathbf{r}}_k)) \leq \pi/2.$$

If you define a new function based on Eq.[B4] as

$$f(r) = \frac{1}{\frac{1}{r} + \frac{1}{\sqrt{4-r^2}}} , \qquad\qquad\qquad [B7]$$

it should be clear that $f$ is concave on $[0,2]$ because $f$ is continuous and twice-differentiable,

and $\frac{d^2 f}{dr^2} \le 0$ for any $r \in (0,2)$. Further, $f(0) = 0$. Therefore, $f$ is sub-additive. We should note that $f$ is an increasing function on $[0,\sqrt{2}]$ and a decreasing function on $[\sqrt{2},2]$.

To prove the triangle inequality, we invoke Lemma 2.2 to focus on one of the triplets, i.e., $\widetilde{\mathbf{r}}_i, \widetilde{\mathbf{r}}_j$,

and $\widetilde{\mathbf{r}}_k$, and write

$$\left\| \widetilde{\mathbf{r}}_i - \widetilde{\mathbf{r}}_j \right\| = a + b \le \sqrt{2} \text{ , with } a = \frac{(\widetilde{\mathbf{r}}_i - \widetilde{\mathbf{r}}_j) \cdot (\widetilde{\mathbf{r}}_k - \widetilde{\mathbf{r}}_j)}{\left\| \widetilde{\mathbf{r}}_i - \widetilde{\mathbf{r}}_j \right\|} \ge 0 \text{ , and } b = \frac{(\widetilde{\mathbf{r}}_j - \widetilde{\mathbf{r}}_i) \cdot (\widetilde{\mathbf{r}}_k - \widetilde{\mathbf{r}}_i)}{\left\| \widetilde{\mathbf{r}}_i - \widetilde{\mathbf{r}}_j \right\|} \ge 0 \text{ .}$$

By Lemma 2.1, $f\left( \left\| \widetilde{\mathbf{r}}_i - \widetilde{\mathbf{r}}_j \right\| \right) = f(a+b) \le f(a) + f(b)$ .

By Lemma 2.2 again, we have

$\frac{(\widetilde{\mathbf{r}}_i - \widetilde{\mathbf{r}}_j) \cdot (\widetilde{\mathbf{r}}_k - \widetilde{\mathbf{r}}_j)}{\left\| \widetilde{\mathbf{r}}_i - \widetilde{\mathbf{r}}_j \right\| \left\| \widetilde{\mathbf{r}}_k - \widetilde{\mathbf{r}}_j \right\|} \le 1$ , and $\frac{(\widetilde{\mathbf{r}}_j - \widetilde{\mathbf{r}}_i) \cdot (\widetilde{\mathbf{r}}_k - \widetilde{\mathbf{r}}_i)}{\left\| \widetilde{\mathbf{r}}_i - \widetilde{\mathbf{r}}_j \right\| \left\| \widetilde{\mathbf{r}}_k - \widetilde{\mathbf{r}}_i \right\|} \le 1$ , and therefore,

$a \le \left\| \widetilde{\mathbf{r}}_k - \widetilde{\mathbf{r}}_j \right\| \le \sqrt{2}$ and $b \le \left\| \widetilde{\mathbf{r}}_k - \widetilde{\mathbf{r}}_i \right\| \le \sqrt{2}$ .

Note that

$$f(a) + f(b) \le f\left( \left\| \widetilde{\mathbf{r}}_k - \widetilde{\mathbf{r}}_j \right\| \right) + f\left( \left\| \widetilde{\mathbf{r}}_k - \widetilde{\mathbf{r}}_i \right\| \right)$$

because $f$ is an increasing function on $[0,\sqrt{2}]$. Therefore,

$$f\left( \left\| \widetilde{\mathbf{r}}_i - \widetilde{\mathbf{r}}_j \right\| \right) \le f\left( \left\| \widetilde{\mathbf{r}}_k - \widetilde{\mathbf{r}}_j \right\| \right) + f\left( \left\| \widetilde{\mathbf{r}}_k - \widetilde{\mathbf{r}}_i \right\| \right).$$

We conclude that $m(\cdot,\cdot)$ satisfies the triangular inequality and therefore, it is a pseudometric.

FIGURE CAPTIONS

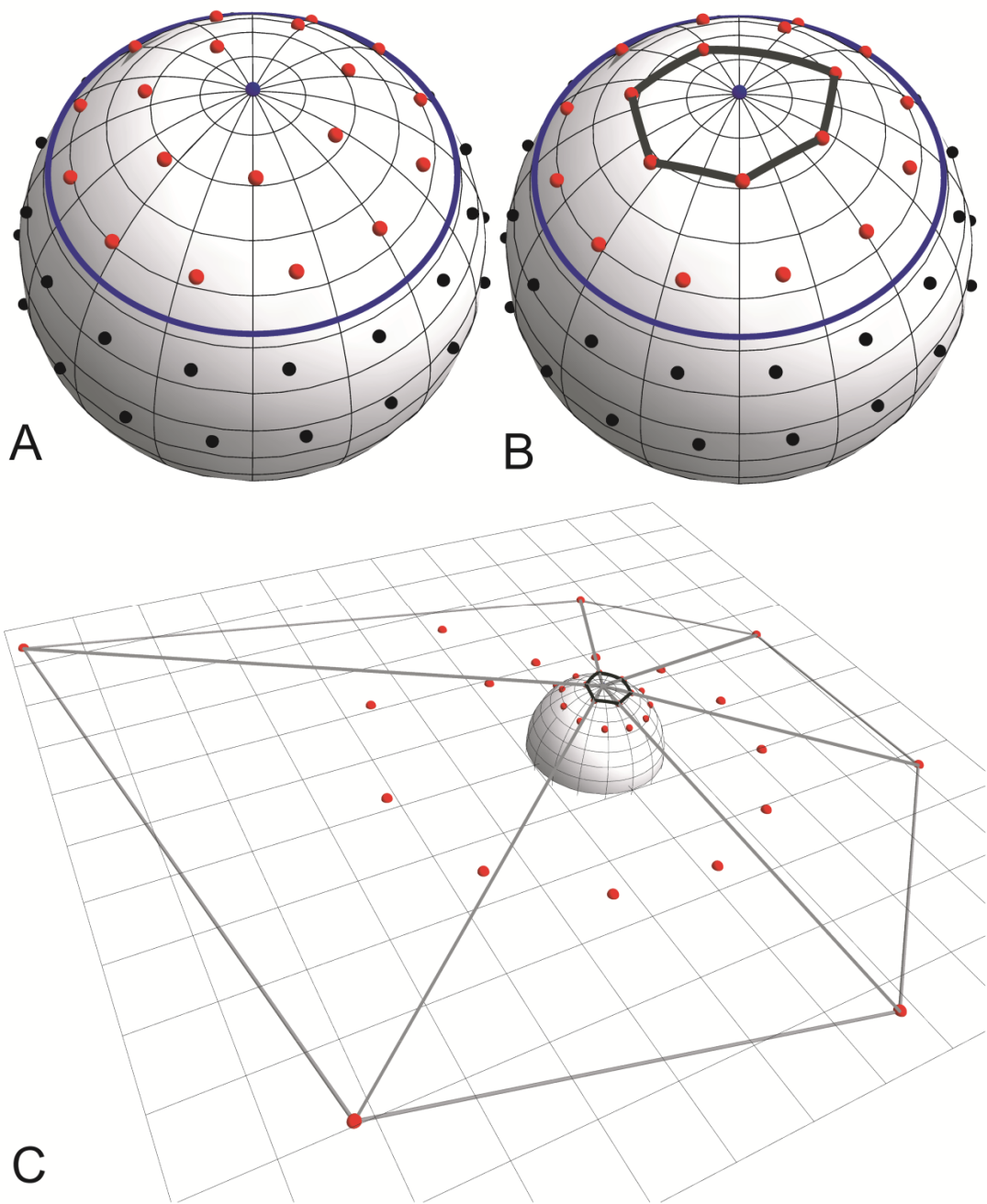

Figure 1. (A) A generator (in blue) and its surrounding generators (in red) within a spherical cap (blue closed curve) of predetermined radius. (B) The smallest convex region formed by the surrounding generators that contains the generator. (C) Finding the smallest convex region that contains the generator is equivalent to finding the convex hull of the surrounding generators in the stereographic projection.

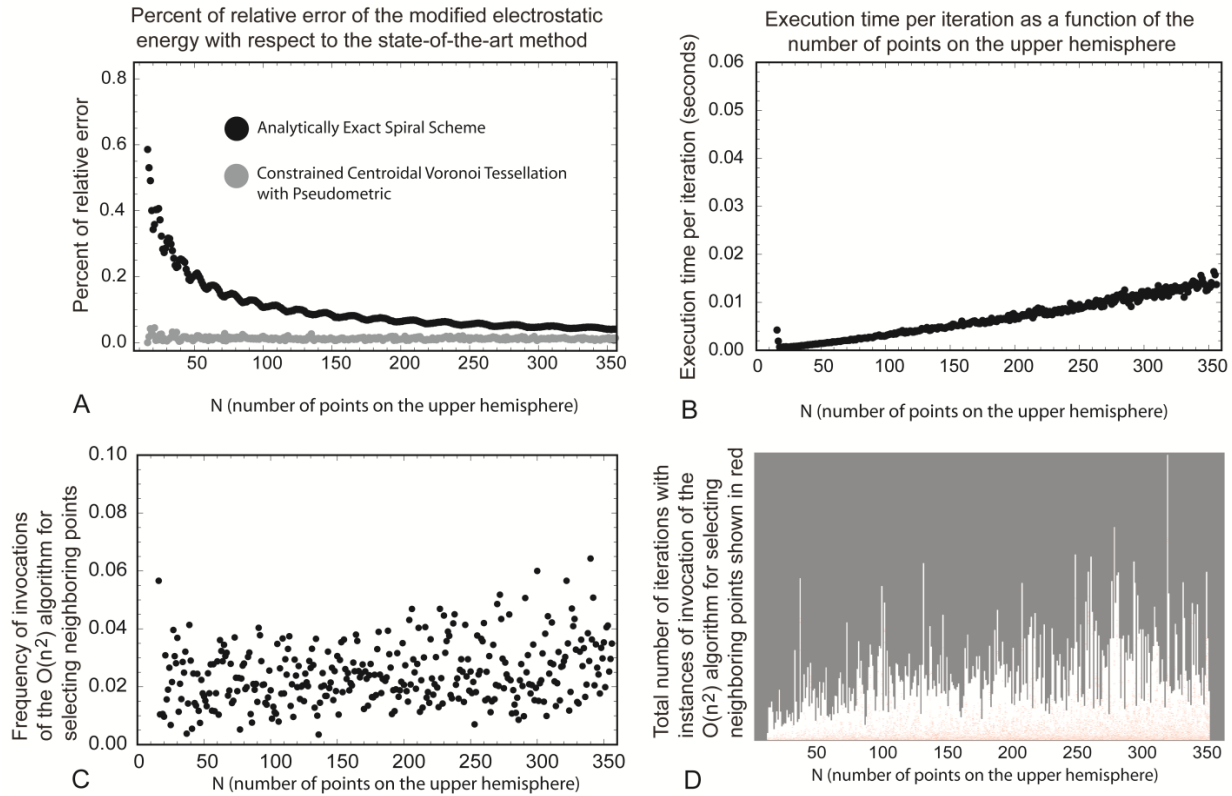

Figure 2. (A) Percent of relative error of initial generators obtained from the analytically exact spiral scheme and of the final generators or centroids obtained from the proposed method compared to the state-of-the-art method in terms of modified electrostatic energies of N antipodally symmetric point sets with N ranges from 16 to 356. (B) The performance of the proposed method in terms of execution time (in seconds) per iteration of the proposed method as a function of the number of points on the upper hemisphere has a linear trend with respect to N. (C) The frequency of invocation of Step 1, which is an $O(N^2)$ algorithm, as a function of N, number of points on the upper hemisphere. (D) Total number of iterations as a function of N with instances of invocation of Step1 throughout the iteration are shown in red. The highest peak is at (N=324, 4161 iterations).

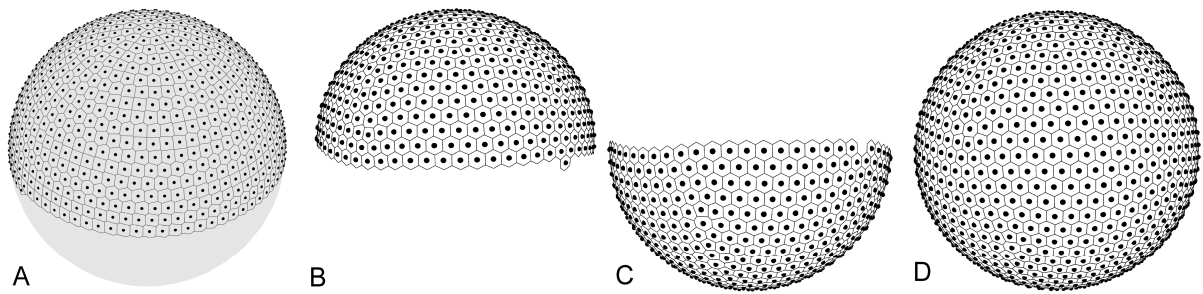

Figure 3. (A) 888 deterministic generators taken from the analytically exact spiral scheme and their Voronoi regions on the upper hemisphere. 888 final (B) real and (C) virtual generators from the proposed method. (D) 1776 antipodally symmetric points on the sphere obtained from (B) and (C).

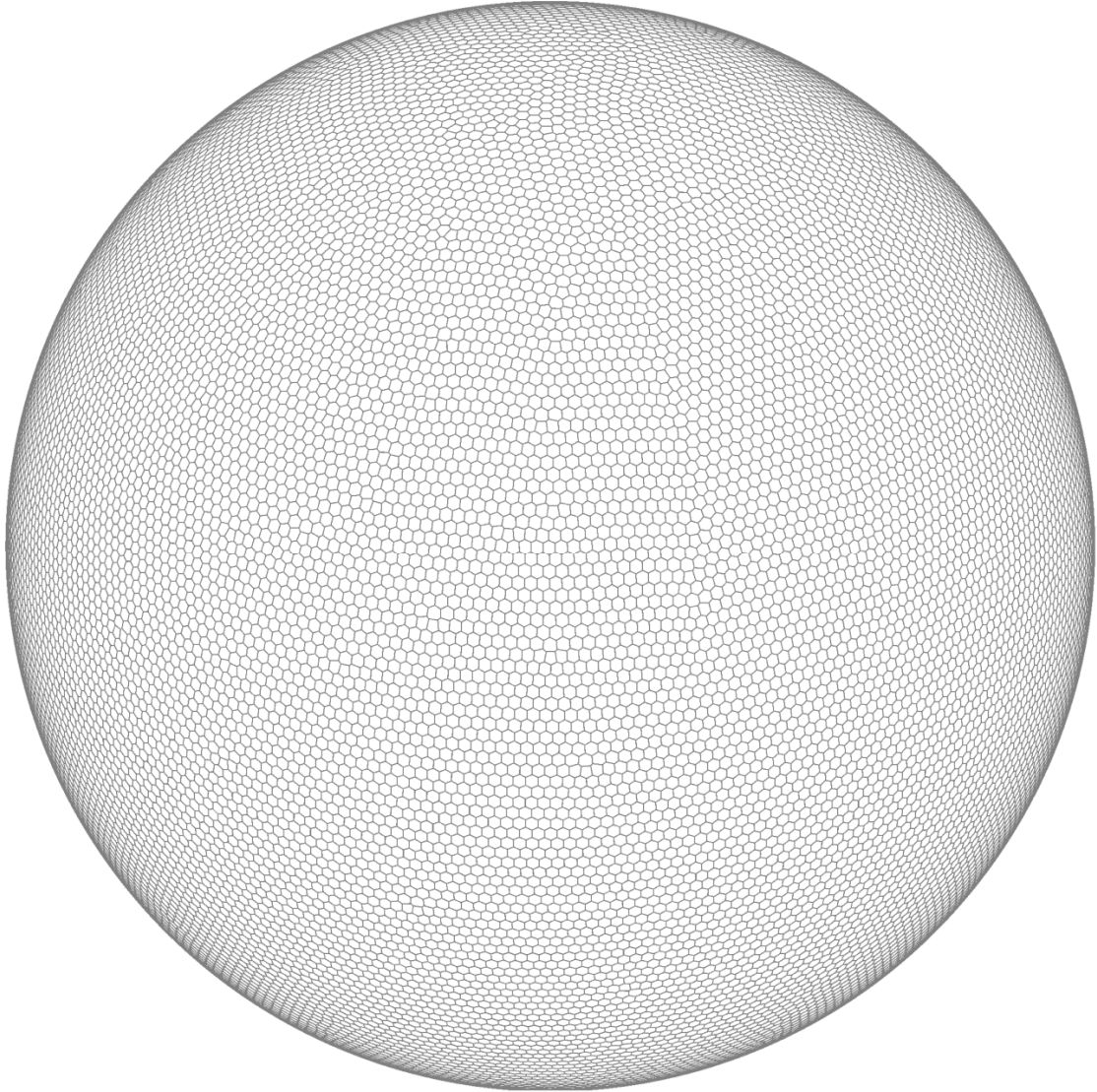

Figure 4. Pseudometrically constrained Voronoi tessellations of the sphere with 16000 generators (not shown in the figure) on the upper hemisphere.

| | The proposed method | Wong and Roos | Generalized Spiral Scheme | Tessellated Icosahedral of the upper hemisphere |
|---|---|---|---|---|
| **Standard Resolution (128x128x128)** z = - 0.181 | 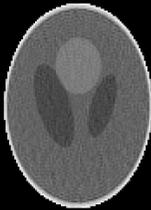 | 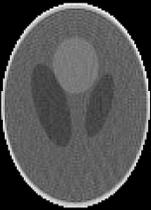 | 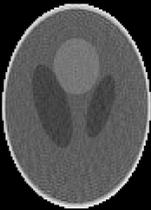 | 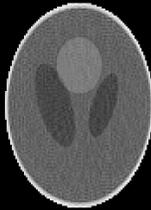 |
| z = 0.228 | 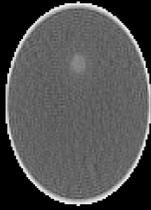 | 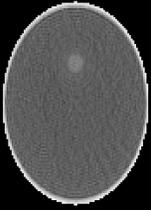 | 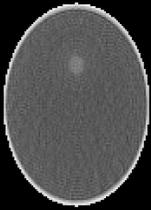 | 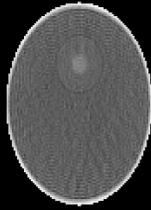 |
| **High Resolution 256x256x256)** z = - 0.239 | 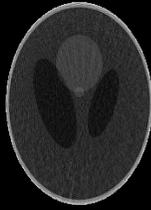 | 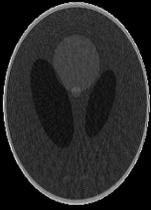 | 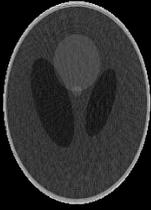 | 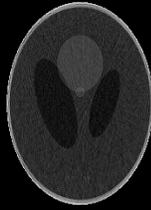 |
| z = 0.051 | 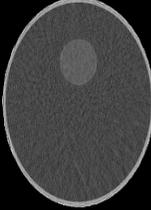 | 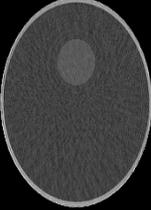 | 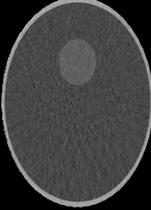 | 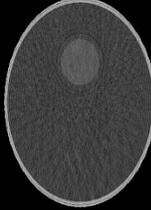 |

Figure 5. Reconstructed images of the analytical phantom generated from different sampling schemes at various slice locations. At the standard resolution of 128x128x128 with z = -0.181, the images generated by the method of Wong and Roos and the generalized spiral scheme have more noticeable ringing artifacts around the bright region than that of the proposed method. The image generated by the tessellated icosahedral scheme has more streak artifacts in the dark regions than that of the proposed method. At the same resolution with z = 0.228, the images generated from these methods have more ringing artifacts than that of the proposed method. Similar patterns of artifacts showed up more noticeably at the high resolution of 256x256x256. At this resolution, the images generated from the proposed method have less ringing artifacts than those from other methods.